\documentclass[fleqn,twoside]{article}
\usepackage{espcrc2}
\usepackage{graphicx}
\usepackage[figuresright]{rotating}
\usepackage{amsmath}

\title{Probing Extra Dimensions with Neutrino Oscillations}

\author{P. A. N. Machado\address[IFUSP]{Instituto de F\'{\i}sica, 
        Universidade de S\~ao Paulo, \\ 
        C.\ P.\ 66.318, 05315-970 S\~ao Paulo, Brazil},\,
        H. Nunokawa\address[PUC]{Departamento de F\'{\i}sica, 
        Pontif\'{\i}cia Universidade Cat\'olica 
        do Rio de Janeiro,\\
        C.\ P.\ 38071, 22452-970 Rio de Janeiro, Brazil} \,
        and
        R. Zukanovich Funchal\addressmark[IFUSP]
        \thanks{Talk given at NOW2010, Neutrino Oscillation Workshop, 
Conca Specchiulla, 
Otranto, Italy, September 4-11, 2010. 
       E-mail: zukanov@fma.if.usp.br}}
       
\begin{document}

\begin{abstract}
We consider a model where sterile neutrinos can propagate in a large
compactified extra dimension ($a$) giving rise to Kaluza-Klein (KK)
modes and the Standard Model left-handed neutrinos are confined to a
4-dimensional spacetime brane.  The KK modes mix with the standard
neutrinos modifying their oscillation pattern. We examine current
experiments in this framework obtaining stringent limits on $a$.
\vspace{1pc}
\end{abstract}

\maketitle
\section{Introduction}
The introduction of singlet neutrino fields which can propagate in 
extra spatial dimensions as well as in the usual three dimensional space 
may lead to naturally small Dirac neutrino masses, due to a volume 
suppression. If those singlets mix with standard neutrinos they may
have an impact on neutrino oscillations,  even if 
the size of the largest extra dimension is smaller than 
$2 \times 10^{-4}$ m (the current limit from Cavendish-type 
experiments which test the Newton Law).

\section{Theoretical Framework}

Here we consider the model discussed in Ref.~\cite{DLP02} where the 3
standard model (SM) left-handed neutrinos $\nu^{\alpha}_{L}$
and the other SM fields, including the Higgs ($H$), are confined to
propagate in a 4-dimensional spacetime, while 3 families of SM singlet
fermions ($\Psi^\alpha$) can propagate in a higher dimensional
spacetime with at least two compactified extra dimensions, one
of these ($y$) compactified on a circle of radius $a$, much larger
than the size of the others so that we can in practice use a
5-dimensional treatment.

The singlet fermions have Yukawa couplings $\lambda_{\alpha \beta}$
with the Higgs and the SM neutrinos leading to Dirac masses and
mixings among active species and sterile KK modes.  This can be
derived from the action
\vglue -0.7cm
\begin{eqnarray*}
S \, &= &\, \int d^4 x \, dy \, \imath \, \Psi^{\alpha} \, \Gamma_{J} \, \partial^{J}\Psi^\alpha \\
&+& \int d^4 x \, \imath \, \bar \nu^{\alpha}_{L} \, \gamma_{\mu} \, \partial^{\mu}
\nu^\alpha_L \\
&+& \int d^4 x \, \lambda_{\alpha \beta} \, H \, \bar \nu^{\alpha}_{L} \, \Psi^\beta_{R} (x,0)+\mbox{h.c.}, 
\end{eqnarray*}
\vglue -0.1cm
where $\Gamma_J, J=0,..,4$ are the 5-dimensional Dirac matrices, that after
dimensional reduction and electroweak symmetry breaking gives rise to
the effective neutrino mass Lagrangian
\vglue -0.6cm
\begin{eqnarray*}
\mathcal{L_{\rm eff}} \, &= & \displaystyle
\sum_{\alpha,\beta}m_{\alpha\beta}^{D}\left[\overline{\nu}_{\alpha
    L}^{\left(0\right)}\,\nu_{\beta R}^{\left(0\right)}+\sqrt{2}\,
  \sum_{N=1}^{\infty}\overline{\nu}_{\alpha
    L}^{\left(0\right)}\,\nu_{\beta
    R}^{\left(N\right)}\right] \\
& + &\sum_{\alpha}\sum_{N=1}^{\infty}\displaystyle
\frac{N}{a}\, \overline{\nu}_{\alpha L}^{\left(N\right)} \,
\nu_{\alpha R}^{\left(N\right)} +\mbox{h.c.}, 
\end{eqnarray*}
\vglue -0.1cm
where the Greek indices $\alpha,\beta = e,\mu,\tau$, the capital Roman
index $N=1,...,\infty$, $m_{\alpha \beta}^{D}$ is a Dirac mass
matrix, $\nu^{(0)}_{\alpha R}$ , $\nu^{(N)}_{\alpha R}$ and
$\nu^{(N)}_{\alpha L}$ are the linear combinations of the singlet fermions 
that couple to the SM neutrinos $\nu^{(0)}_{\alpha L}$.

In this context one can compute the active neutrino transition probabilities
\[  P(\nu_{\alpha}^{(0)}\to\nu_{\beta}^{(0)};L)= \] \vglue -0.9cm
\[   \bigg|\sum_{i,j,k}\sum_{N=0}^{\infty}U_{\alpha i}U_{\beta k}^{*}W_{ij}^{(0N)*}W_{kj}^{(0N)}\exp\left(i\frac{\lambda_{j}^{(N)2}L}{2Ea^{2}}\right)\bigg|^{2}
\]
where $U$ and $W$ are the mixing matrices for active and KK modes,
respectively. Here $\lambda_j^{(N)}$ is a dimensionless eigenvalue of
the evolution equation~\cite{DLP02} which depends on the $j$-th
neutrino mass ($m_j$), hence on the mass hierarchy, $L$ is the baseline and $E$
is the neutrino energy.

\begin{figure}[htb]
\begin{center}
\includegraphics[width=0.50\textwidth]{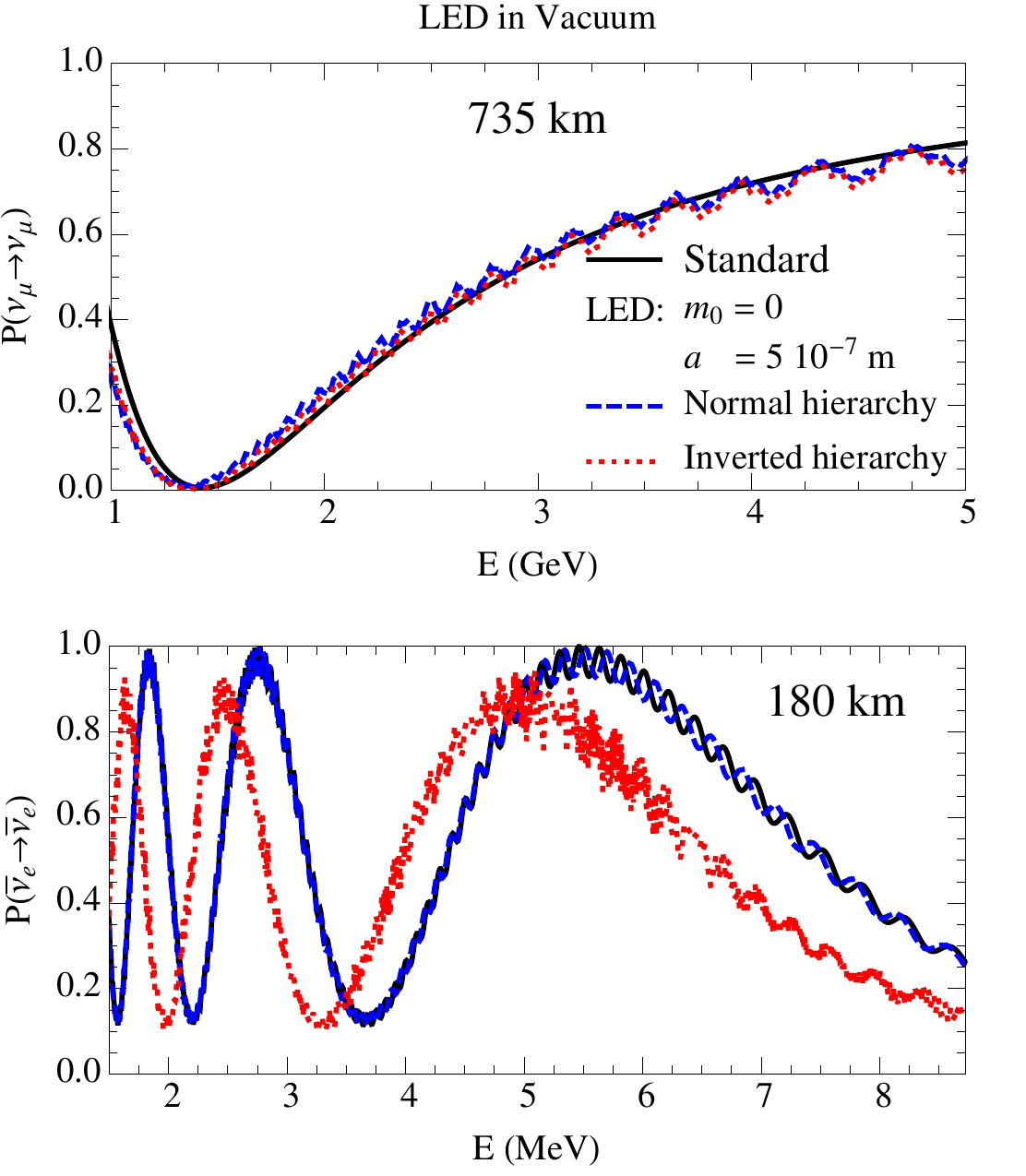}
\end{center}
\vglue -1.3cm
\caption{In the top (bottom) panel we show the survival probability for 
$\nu_\mu$ ($\bar \nu_e$) as a function of the neutrino energy for 
$L=735$ km ($180$ km) for $a=0$ (no LED, black curve) and  
$a=5\times 10^{-7}$ m for normal hierarchy (dashed blue curve) and 
inverted hierarchy (dotted red curve).}
\label{fig:prob}
\vglue -0.7cm
\end{figure}

To illustrate what is expected we plot in Fig.~\ref{fig:prob} the survival 
probabilities  for $\nu_\mu \to \nu_\mu$ and $\bar \nu_e \to \bar \nu_e$ as 
a function of $E$. We show the behavior for the normal and inverted mass 
hierarchy assuming the lightest neutrino to be massless ($m_0=0$).
The effect of this large extra dimension (LED) depends on the product $m_j a$.
We observe that in the $\nu_\mu \to \nu_\mu$ channel the effect of LED 
is basically the same for normal (NH) and inverted (IH) hierarchies, since in 
this case all the amplitudes involved are rather large.
On the other hand, for $\bar \nu_e \to \bar \nu_e$ the effect is smaller for 
NH as in this case the dominant $m_j a$ term is suppressed by $\theta_{13}$.

\section{Results}
As we can observe in Fig.~\ref{fig:prob} the main effect of LED is a shift in 
the oscillation maximum with a decrease in the survival probability due to 
oscillations to KK modes. This makes experiments such as KamLAND and MINOS, 
which are currently the best experiments to measure $\Delta m^2_{\odot}$ and  
$\vert \Delta m^2_{\rm atm}\vert$, respectively, also the best experiments to 
test for LED.

We have used the recent MINOS~\cite{MINOS10} and KamLAND~\cite{Kam08}
results and reproduced their allowed regions for the standard
oscillation parameters. For this and the LED study we have modified
GLoBES \cite{globes} according to our previous analysis of these experiments 
in \cite{MNTZ05} and \cite{MKP10}.

\begin{figure}[!h]
\begin{center}
\includegraphics[width=0.42\textwidth]{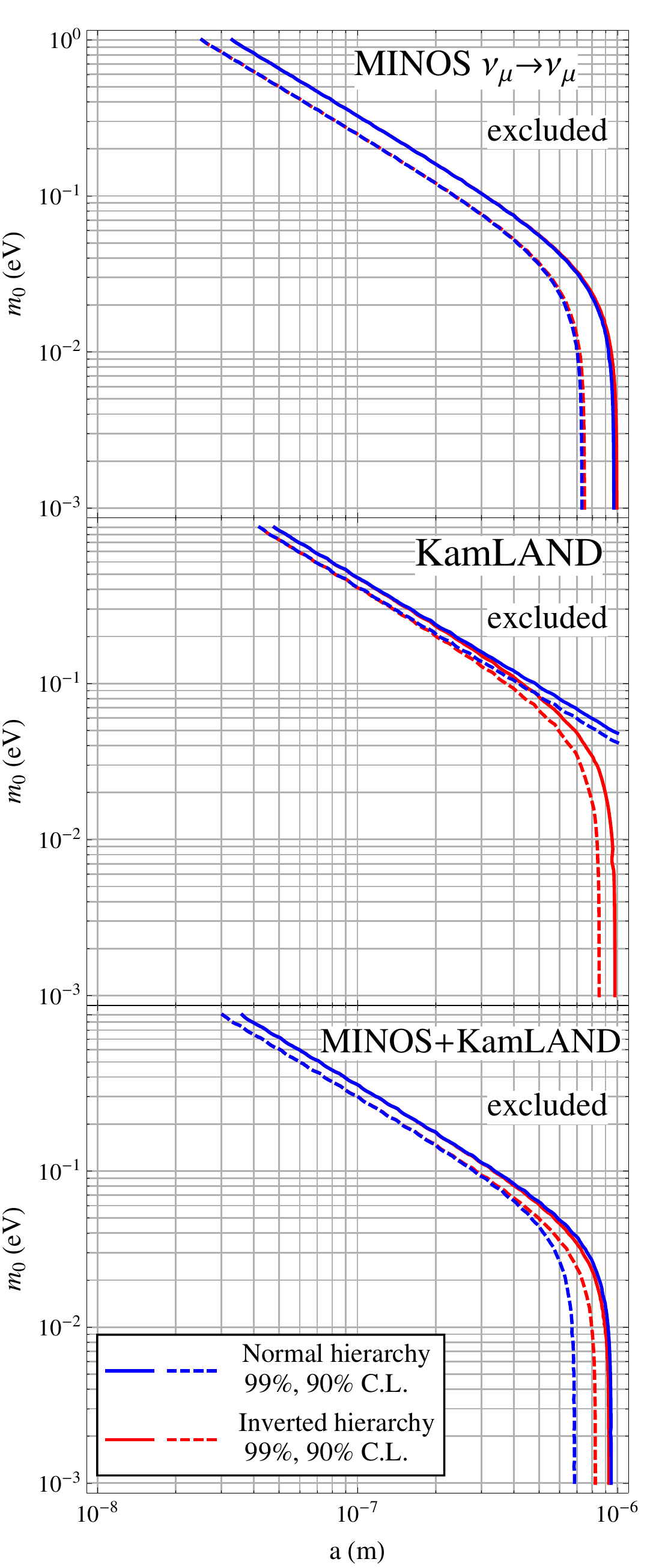}
\end{center}
\vglue -1.3cm
\caption{Excluded region in the plane $m_0 \times a$ by KamLAND (upper panel),
 MINOS (center panel) and combined (lower panel).
}
\label{fig:exc}
\vglue -0.5cm
\end{figure}

In Fig.~\ref{fig:exc}, we present the excluded region in the plane $m_0
\times a$, by MINOS, KamLAND and their combined data at 90 and 99\%
CL (2 dof). When finding these regions all standard oscillation parameters
where considered free. To account for our previous knowledge of their 
values \cite{concha}, we have added Gaussian priors to the $\chi^2$ function.
As expected the limits provided by MINOS ($\nu_\mu \to \nu_\mu$) are 
basically the same for NH and IH. From their data we obtain  
$a < 7.3 (9.7) \times 10^{-7}$ m in the hierarchical 
case for $m_0 \to 0$ and $a < 1.2 (1.6) \times 10^{-7}$ m at 90 (99)\%~CL 
for degenerate neutrinos with $m_0 = 0.2$ eV. We have verified that the 
inclusion of LED in the fit of the standard atmospheric oscillation 
parameters does not modify very much the region in the plane 
$\sin^2 2\theta_{23} \times \vert \Delta m^2_{\rm atm} \vert$   allowed 
by MINOS data. In fact the best fit point as well as the $\chi^2_{\rm min}$ 
remain the same as in the case without LED.

KamLAND data provide, for hierarchical neutrinos with $m_0 \to 0$, a
competitive limit only for IH, in this case one gets $a < 8.5 (9.8)
\times 10^{-7}$ m at 90 (99)\% CL. For degenerate neutrinos with
$m_0= 0.2$ eV one also gets from KamLAND $a < 2.0 (2.3) \times
10^{-7}$ m at 90 (99)\% CL. The inclusion of LED in the fit of the
standard solar oscillation parameters here enlarges the region in the
plane $\tan^2 \theta_{12} \times \Delta m^2_{\odot}$ allowed by
KamLAND data. The best fit point changes from $\Delta m^2_{\odot} =
7.6 \times 10^{-5}$ eV$^2$ and $\tan^2 \theta_{12}=0.62$ to $\Delta
m^2_{\odot} = 8.6 \times 10^{-5}$ eV$^2$ and $\tan^2 \theta_{12}=0.42$,
however, the $\chi^2_{\rm min}/\rm dof$ remains the same.

When one combines both experiments the hierarchical limits 
improve but the degenerate limit remains practically the one given by MINOS.

\section{Conclusions}
\vglue -0.1cm
We have investigated the effect of LED in neutrino oscillation data
deriving limits on the largest extra dimension $a$ provided by the
most recent data from MINOS and KamLAND experiments. For hierarchical
neutrinos with $m_0 \to 0$, MINOS and KamLAND constrain $a < 6.8 (9.5)
\times 10^{-7}$ m for NH and $a < 8.5 (9.8) \times 10^{-7}$ m 
at 90 (99)\% CL for IH.  For degenerate neutrinos with
$m_0= 0.2$ eV their combined data constrain $a < 2.1 (2.3) \times
10^{-7}$ m at 90 (99)\% CL. 

We can estimate that the future Double CHOOZ experiment will be able to 
improve these limits by a factor 2 for the IH and by a factor 1.5 for 
the degenerate case. Unfortunately NO$\nu$A and T2K cannot improve 
MINOS limits. See~\cite{MNZ-LED} for any detail.

RZF and HN thank Profs. Fogli and Lisi for the cordial invitation
to participate of NOW2010. This work has been 
        supported by FAPESP, FAPERJ and CNPq Brazilian 
funding agencies.

\end{document}